\documentclass[12pt,preprint]{aastex}

\slugcomment{}

\shorttitle{X-Ray Flash XRF 020903}
\shortauthors{Sakamoto et al.}

\begin{document}


\title{HETE-2 Observations of the Extremely Soft X-Ray Flash XRF 020903}


\author{T.Sakamoto\altaffilmark{1,2,3}, 
D. Q. Lamb\altaffilmark{4}, 
C. Graziani\altaffilmark{4},
T. Q. Donaghy\altaffilmark{4}, 
M. Suzuki\altaffilmark{1},  
G. Ricker\altaffilmark{5},
J-L. Atteia\altaffilmark{6}, 
N. Kawai\altaffilmark{1,2}, 
A. Yoshida\altaffilmark{2,7}, 
Y. Shirasaki\altaffilmark{8}, 
T. Tamagawa\altaffilmark{2}, 
K. Torii\altaffilmark{2}, 
M. Matsuoka\altaffilmark{9}, 
E. E. Fenimore\altaffilmark{3}, 
M. Galassi\altaffilmark{3}, 
T. Tavenner\altaffilmark{10},  
J. Doty\altaffilmark{5},
R. Vanderspek\altaffilmark{5},
G. B. Crew\altaffilmark{5},
J. Villasenor\altaffilmark{5},
N. Butler\altaffilmark{5}, 
G. Prigozhin\altaffilmark{5},
J. G. Jernigan\altaffilmark{11},
C. Barraud\altaffilmark{6},
M. Boer\altaffilmark{12}, 
J-P. Dezalay\altaffilmark{12}, 
J-F. Olive\altaffilmark{12}, 
K. Hurley\altaffilmark{11}, 
A. Levine\altaffilmark{5},
G. Monnelly\altaffilmark{5},
F. Martel\altaffilmark{5},
E. Morgan\altaffilmark{5},
S. E. Woosley\altaffilmark{13},
T. Cline\altaffilmark{14},
J. Braga\altaffilmark{15},
R. Manchanda\altaffilmark{16},
G. Pizzichini\altaffilmark{17},
K. Takagishi\altaffilmark{18},
and M. Yamauchi\altaffilmark{18}} 

\altaffiltext{1}{Department of Physics, Tokyo Institute of Technology,
2-12-1 Ookayama, Meguro-ku, Tokyo 152-8551, Japan}
\altaffiltext{2}{RIKEN (Institute of Physical and Chemical Research),
2-1 Hirosawa, Wako, Saitama 351-0198, Japan}
\altaffiltext{3}{Los Alamos National Laboratory, P.O. Box 1663, Los
Alamos, NM, 87545}
\altaffiltext{4}{Department of Astronomy and Astrophysics, University
of Chicago, IL, 60637}
\altaffiltext{5}{Center for Space Research, Massachusetts Institute
of Technology, 70 Vassar Street, Cambridge, MA, 02139}
\altaffiltext{6}{Laboratoire d'Astrophysique, Observatoire
Midi-Pyre\'ne\'es, 14 Ave. E. Belin, 31400 Toulouse, France}
\altaffiltext{7}{Department of Physics, Aoyama Gakuin University,
Chitosedai 6-16-1, Setagaya-ku, Tokyo 157-8572, Japan}
\altaffiltext{8}{National Astronomical Observatory, Osawa 2-21-1,
Mitaka, Tokyo 181-8588, Japan}
\altaffiltext{9}{Tsukuba Space Center, National Space Development
Agency of Japan, Tsukuba, Ibaraki, 305-8505, Japan}
\altaffiltext{10}{Department of Astronomy, New Mexico State University,
1320 Frenger Mall, Las Cruces, NM, 88003-8001}
\altaffiltext{11}{University of California at Berkeley, Space Sciences
Laboratory, Berkeley, CA, 94720-7450}
\altaffiltext{12}{Centre d'Etude Spatiale des Rayonnements, CNRS/UPS,
B.P.4346, 31028 Toulouse Cedex 4, France}
\altaffiltext{13}{Department of Astronomy and Astrophysics, University
of California at Santa Cruz, 477 Clark Kerr Hall, Santa Curz, CA 95064}
\altaffiltext{14}{NASA Goddard Space Flight Center, Greenbelt, MD, 20771}
\altaffiltext{15}{Instituto Nacional de Pesquisas Espaciais, Avenida
Dos Astronautas 1758, Sa\~o Jose\' dos Campos 12227-010, Brazil}
\altaffiltext{16}{Department of Astronomy and Astrophysics, Tata
Institute of Fundamental Research, Homi Bhabha Road, Mumbai, 400 005, India}
\altaffiltext{17}{Consiglio Nazionale delle Ricerche (IASF), via Piero
Gobetti, 101-40129 Bologna, Italy}
\altaffiltext{18}{Faculty of engineering, Miyazaki University, Gakuen
Kibanadai Nishi, Miyazaki 889-2192, Japan}




\begin{abstract} 
We report HETE-2 WXM/FREGATE observations of the X-ray flash, 
XRF 020903.  This
event was extremely soft: the ratio $\log(S_X/S_\gamma) = 0.7$, where
$S_X$ and $S_\gamma$ are the fluences in the 2-30 and 30-400 keV energy
bands, is the most extreme value observed so far by HETE-2.  In
addition, the spectrum has an observed peak energy $E^{\rm obs}_{\rm
peak} < 5.0$ keV (99.7\% probability upper limit) and no photons were
detected above $\sim$ 10 keV.  The burst is shorter at higher energies,
which is similar to the behavior of long GRBs.  We consider the
possibility that the burst lies at very high redshift and that the low
value of $E_{\rm peak}^{\rm obs}$ is due to the cosmological redshift,
and show that this is very unlikely.  We find that the properties of
XRF 020903 are consistent with the relation between the fluences
$S(7-30~{\rm keV})$ and $S(30-400~{\rm keV})$ found by Barraud et al.
for GRBs and X-ray-rich GRBs, and are consistent with the extension by
a decade of the hardness-intensity correlation (Mallozzi et al. 1995)
found by the same authors.  Assuming that XRF 020903 lies at a redshift
$z = 0.25$ as implied by the host galaxy of the candidate optical and
radio afterglows of this burst, we find that the properties of XRF
020903 are consistent with an extension by a factor $\sim 300$ of the
relation between the isotropic-equivalent energy $E_{\rm iso}$ and the
peak $E_{\rm peak}$ of the $\nu F_\nu$ spectrum (in the source frame of
the burst) found by Amati et al. for GRBs.  The results presented in
this paper therefore provide evidence that XRFs, X-ray-rich GRBs, and
GRBs form a continuum and are a single phenomenon.  The results also
impose strong constraints on models of XRFs and X-ray-rich GRBs.
\end{abstract}




\section{Introduction}

Gamma-ray bursts (GRBs) that have a large fluence in the X-ray energy
band (2-30 keV) relative to the gamma-ray energy band (30-400 keV) are
receiving increased attention.  The Burst and Transient Source 
Experiment (BATSE) on board the {\it Compton Gamma Ray Observatory}
detected 2704 GRBs \citep{paciesas1999}.  The spectra of 156 bright
bursts exhibit a distribution of low-energy power-law indices $\alpha$
whose centroid is $\sim -1$, and a distribution of observed break
energies $E_{\rm break}^{\rm obs}$ whose centroid is $\approx$ 230 keV
\citep{preece2000}, where $E_{\rm break}^{\rm obs} = (\alpha - \beta) (2
+ \alpha)^{-1} E_{\rm peak}^{\rm obs}$.  Here $\alpha$, $\beta$, and
$E_{\rm peak}^{\rm obs}$ are the slope of the low-energy power-law
index, the high-energy power-law index, and the energy of the peak of
the $\nu F_\nu$ spectrum of the Band function \citep{band1993}, an
expression that satisfactorily represents the spectra of almost all
GRBs.  In contrast, 36\% of the bright bursts observed by {\it GINGA}
have peak energies $E_{\rm peak}^{\rm obs}$ in their photon number
spectrum at a few keV and large X-ray to $\gamma$-ray fluence ratios
\citep{strohmayer1998}.  

The {\it Beppo}SAX Wide Field Camera (WFC) detected events that are
very similar to the soft {\it GINGA} GRBs; these events have been
termed ``X-ray Flashes'' (XRFs) \citep{heise2000}.\footnote{Throughout
this paper, we define ``X-ray-rich'' GRBs and XRFs as those events for
which $\log [S_X(2-30~{\rm kev})/S_\gamma(30-400~{\rm kev})] > -0.5$
and 0.0, respectively.}  The energy flux of these XRFs lies in the
range 10$^{-8} - 10^{-7}$ erg cm$^{-2}$ s$^{-1}$ and the low-energy
photon index $\alpha$ of their spectra ranges from $-$3 to $-$1.2.  The
sky distribution of XRFs is consistent with isotropy, and there is no
evidence that the sources are Galactic.  The XRFs have t$_{90}$
durations between 10 and 200 sec.  The event rate of XRFs detected
 by the WFC is
3.4  events per year.  Clarifying the connection between XRFs and GRBs 
could provide a breakthrough in our understanding of the prompt
emission of GRBs. 

\cite{kippen2002} made a detailed spectral comparison of GRBs and
XRFs,  using a sample of eighteen GRBs that were observed by BATSE and
a sample of nine  XRFs that were observed by both the WFC and BATSE.  
According to their joint analysis of WFC/BATSE spectral data, the
low-energy and high-energy photon indices of XRFs are $-1$ and $\sim
-2.5$, respectively, which are no different from those of GRBs.  On the
other hand, XRFs have much lower values of $E_{\rm peak}^{\rm obs}$
than do GRBs.  Thus the only temporal or spectral difference between
GRBs and XRFs appears to be that XRFs have lower $E_{\rm peak}^{\rm
obs}$ values.  Kippen et al. therefore suggest that XRFs might
represent an extension of the GRB population to events with low peak
energies.  Analyzing 35 HETE-2 GRBs seen by FREGATE,
\citet{celine2003} demonstrate that the spectral properties of
``X-ray rich'' GRBs form a continuum with those of ordinary GRBs and
suggest that XRFs may represent a further extension of this continuum.

BATSE's low-energy threshold of $\sim$ 20 keV made it  difficult for
BATSE to detect XRFs.  ${\it Ginga}$ and ${\it Beppo}$SAX had the
capability of detecting XRFs; however, ${\it Ginga}$ could not
determine the direction of the burst and the ${\it Beppo}$SAX GRBM had
difficulty in triggering on XRFs.  Consequently, these missions could
not carry out in depth investigations of XRFs.  In contrast, HETE-2 
\citep{ricker2003} has the ability to trigger on and localize XRFs, and
to study their spectral properties, using the Wide-Field X-Ray Monitor
[WXM; 2-25 keV energy band; \citet{kawai2003}] and the French Gamma
Telescope [FREGATE; 6-400 keV energy band; \citet{atteia2003}], which
have energy thresholds of a few keV.  

In this Letter, we report the detection and localization of XRF 020903
by HETE-2 \citep{ricker2002} and present the results of a detailed
study of its properties.  Since this event was extremely soft and there
was very little signal (a $\sim 2\sigma$ excess in the best selected
energy range) in FREGATE, we focus our analysis on the WXM temporal and
spectral data for the event.

\section{Observations}




\subsection{Localization}

XRF 020903 was detected with the HETE-2 WXM and the Soft X-ray Camera
[SXC; 0.5-10 keV energy band; \cite{villasenor2003}] instruments  at
10:05:37.96 UT on 2002 September 3 \citep{ricker2002}.   The WXM flight
localization was correct, but was not sent out because HETE-2 was
pointing at the Galactic Bulge region at the time and WXM triggers were
therefore not being sent to the GCN in order not to overwhelm the
astronomical community with X-ray burst localizations.  A GCN Notice
reporting the localization of the burst, based on ground analysis 
\citep{graziani2003,shirasaki2003} of the WXM data, was sent out 231
minutes after the burst.  

The WXM localization can be expressed as a 90\% confidence circle that
is  16.6$^{\prime}$ in radius and is centered at 
R.A. = 22$^{\rm h}$49$^{\rm m}$25$^{\rm s}$,   
Dec. = $-$20$^{\circ}$53$^{\prime}$59$^{\prime\prime}$ (J2000).    
A
localization of the burst based on ground analysis \citep{monnelly2003}
of the SXC data was distributed as a GCN Notice about 7 hours after the
burst.  Only a one-dimensional localization was possible using the SXC
data, but this significantly reduced the area of the localization
region for XRF 020903. The improved localization produced by combining
the SXC and WXM localizations can be described as a 90\% confidence
quadrilateral that is 4$^{\prime}$ in width and $\sim$31$^{\prime}$ in
length (see Figure 1).  It is centered at 
R.A. = 22$^{\rm h}$49$^{\rm m}$01$^{\rm s}$, 
Dec. = $-$20$^{\circ}$55$^{\prime}$47$^{\prime\prime}$ (J2000), 
and its four corners lie at 
(R.A., Dec.) = (22$^{\rm h}$48$^{\rm m}$48.00$^{\rm s}$,
$-$20$^{\circ}$39$^{\prime}$36.0$^{\prime\prime}$),
(22$^{\rm h}$48$^{\rm m}$33.60$^{\rm s}$,
$-$20$^{\circ}$42$^{\prime}$36.0$^{\prime\prime}$),
(22$^{\rm h}$49$^{\rm m}$10.80$^{\rm s}$,
$-$21$^{\circ}$10$^{\prime}$12.0$^{\prime\prime}$), and
(22$^{\rm h}$49$^{\rm m}$30.00$^{\rm s}$,
$-$21$^{\circ}$10$^{\prime}$48.0$^{\prime\prime}$) (J2000).  

Detections of candidate optical and radio afterglows of XRF 020903, and
the host galaxy of the candidate optical and radio afterglows, have
been reported.  \cite{soderberg2002} discovered an
optical transient within the HETE-2 SXC + WXM
localization region at  
R.A. = $22^{\rm h}48^{\rm m}42.34^{\rm s}$,  
Dec = $-20^{\circ}46^{\prime}09.3^{\prime\prime}$,
using the Palomar 200-inch telescope.  These authors mention that the
optical transient brightened by $\sim 0.3-0.4$ magnitudes between about
7 and 24 days after the XRF, and suggest that the re-brightening might
be due to an associated supernova.  However, the optical transient 
apparently faded by over a magnitude only three days later
\citep{covino2002}.  Spectroscopic observations of the optical
transient, using the Magellan 6.5m Baade and Clay telescopes, detected
narrow emission lines from an underlying galaxy at a redshift $z = 0.25
\pm 0.01$, suggesting that the host galaxy of the optical transient is
a star-forming galaxy [\cite{soderberg2002}; see also
\cite{chandfili2002}].  A fading bright radio source at the position of
optical transient was detected using the Very Large Array
\citep{berger2002}.  Hubble Space telescope observations of the XRF
020903 field reveal the optical transient and show that its host galaxy
is an irregular galaxy, possibly with four interacting components
\citep{levan2002}.  These detections likely represent the first
discoveries of the optical and radio afterglows, together with the host
galaxy, of an XRF. 

In our analysis of the prompt emission of XRF 020903, we apply a
``cut''  to the WXM photon time- and energy-tagged data (TAG data),
using only the photons from the pixels on the three wires in the
X-detectors (XA0, XA1, and XA2) and the four wires in the Y-detectors
(YA1, YA2; YB0, YB1) that were illumnated by the burst {\it and} that
maximize the S/N of the burst light curve, in the same manner as we did
for GRB 020531 \citep{lamb2003b}.  We use this optimized TAG data when
performing our temporal and spectral analyses of this event.    

\subsection{Temporal Properties}

Figure 2 shows the light curve of XRF 020903 in four WXM energy bands. 
The time history of the burst in the 2-5 and 5-10 keV energy bands has
two peaks. Clearly, there is no significant flux above 10 keV. 
Table 1 gives the t$_{50}$ and t$_{90}$ durations of the burst in the
2-5 keV, 5-10 keV, and 2-10 keV energy bands.  The duration of the
burst is longer in the lower energy band; this trend is similar to
that seen in long bright GRBs \citep{fenimore1995}.

\subsection{Spectrum}

As we have seen, the light curve of XRF 020903 shows two peaks: the
first occurring in the time interval 0-8 s, and the second occurring in
the time interval 8-13 s.  The S/N of the first peak is much higher
than that of the second.  In addition, inspection of the burst light
curve in the 2-5 and 5-10 keV energy bands suggests that the second
peak is much softer than the first.  For these reasons, we  analyze the
spectrum  of the burst in three time intervals: 0-8~s, 8-13~s, and the
total duration of the burst, 0-13~s.  The background region we use is
40 seconds in duration and starts 45 seconds before the burst.  

The WXM detector response matrix has been well-calibrated using
observations of the Crab nebula \citep{shirasaki2002}.  In the spectral
fits, we include only the photons that registered on the four wires in
the X-detectors and the five wires in the Y-detectors that were
illuminated by the burst, as mentioned above.  Since the variation in
the gain is not uniform at the ends of the wires in the WXM detectors
\citep{shirasaki2000}, we use only the photon counts that registered in
the central $\pm$50 mm region of the wires to construct the spectra of
the burst.  We include all of the photons that register in the central
regions of these wires (i.e, we use the full 2-25 keV energy range of
the WXM).  The relation between pulse height and energy in the WXM is
non-linear and is different for each wire.  In order to extract the
strongest possible constraints on the parameters of the spectral models
we consider, we treat each individual WXM wire separately but take the
normalizations on all wires to be the same.  For the same reason, we do
not re-bin any of the pulse height channels in the WXM and in the
FREGATE, and we carry out a set of fits for the total duration of the
burst (0-13 s) that include  the spectral data from both the WXM and
the FREGATE.  We use the XSPEC v11.2.0 software package to do the
spectral fits.

Table 2 presents the results of our time-resolved and time-integrated
spectral analysis of the burst.  In this analysis, we consider the
following models:  (1) blackbody, (2) power-law model, 
(3) power-law times exponential (PLE) [the
COMP model in \citet{preece2000}], and (4) Band function
\citep{band1993}. 
The Galactic value of $N_H$ in the direction of the burst is $2.3
\times 10^{20}$ cm$^{-2}$ \citep{dickey1990}, which is negligible 
(i.e., it is undetectable
in the WXM energy range).  Furthermore, the WXM and FREGATE data do not
request $N_H$ as a free parameter (i.e., introducing $N_H$ as an
additional free parameter produces only a small change in $\chi^2$). 
There is therefore no need to include $N_H$ as a parameter in the
spectral fits, and we do not do so.  In the Band model fits, we have
fixed $\alpha = -1$, which is a typical value for GRBs, in order to
better constrain the remaining parameters.  All of the models provide
acceptable fits to the data; i.e., the data do not request models more
complicated than a blackbody or a power-law.  

However, essentially all GRB spectra are well-described by the Band
function \citep{band1993,preece2000}, and the analysis of
\cite{kippen2002} shows that at least some XRF spectra are also
well-described by the Band function.  Fits to the WXM data for all 
three time intervals using the power-law model give spectral slopes
$\alpha < -2$ with high significance.  For example, comparing the
minimum value $\chi^2_{\rm min} = 53.4$ corresponding to the best-fit
value of the spectral slope $\alpha = -2.8$ and the value $\chi^2 =
60.6$ at $\alpha = -2$ for the power-law fit to the average spectrum
of the burst (i.e., the time interval 0-13 s), we find that $\Delta
\chi^2 = 7.2$ for one additional parameter.  Thus, using the Maximum
Likelihood Ratio Test, $\alpha < -2$ at the 99.3\% confidence level. 
From this evidence, we conclude that the peak $E_{\rm peak}$ in $\nu
F_\nu$ lies near or below 2 keV, the lower limit of the energy range of
the WXM detectors. 

There is evidence of spectral softening between the first and second
time intervals.  In particular, a power-law fit to the first time
interval gives $\alpha_1  = -2.4^{+0.5}_{-0.6}$ and $\chi^2_{\rm min,1}
= 75.1$, while a power-law fit to the second time interval gives
$\alpha_2 = -4.2^{+1.1}_{-3.7}$ and $\chi^2_{\rm min,2} = 71.1$.  In
contrast, a power-law fit to the first and second time intervals, but
with $\alpha = \alpha_1 = \alpha_2$ gives $\alpha =
-2.86^{+0.44}_{-0.82}$ and $\chi^2_{\rm min} = 152.2$.  The first (more
complicated) model includes the second model as a special case (i.e.,
the models are nested).  Comparing $\chi^2_{\rm min}$ for the two
models, we find that $\Delta \chi^2 = 6.0$ for one additional
parameter.  Thus, using the Maximum Likelihood Ratio Test , there is
evidence of spectral softening at the 98.6\% confidence level.

Figure 3 shows a comparison of the observed count spectrum and the
count spectrum predicted by the best-fit power-law model, for the time
intervals 0-8 s and 8-13s.  Figure 4 shows the same comparison, except
for the total duration of the burst (0-13 s).  Table 3 gives the peak
photon number and energy fluxes (in 1 s) and the fluence of XRF 020903,
assuming the power-law model.\footnote{We compute the peak photon
number flux in the WXM 2-5, 5-10, and 2-10 keV energy bands, using the
best-fit power-law model parameters for the average photon energy
spectrum of the burst, and the ratio 2.731 of the photon flux in the 1 s
time interval containing the largest number of photons and the average
photon flux in the 0 - 13 s time interval.  We compute the peak photon
energy flux in the WXM 2-5, 5-10, and 2-10 keV energy bands in exactly
the same way, except that we use the ratio 3.247 of the total photon
energy flux (found by weighting each photon with its energy and summing
the energies) found in the 1 s time interval containing the largest
total photon energy and the average photon energy flux in the 0 - 13 s
time interval.}  

Using the power-law model that best fits the burst-averaged WXM
plus FREGATE spectral data, 
we find fluences $S_X(2-30~{\rm keV}) = 8.2_{-2.3}^{+2.5} 
\times 10^{-8}$ erg
cm$^{-2}$ and $S_\gamma (30-400~{\rm keV}) = 1.6_{-1.3}^{+4.2} 
\times 10^{-8}$ erg
cm$^{-2}$ where the quoted errors give the 90\% confidence
intervals.  Thus the ratio of fluences $\log[S_X (7-30~{\rm
keV})/S_\gamma (30-400~{\rm keV})] = 0.7$, with a 90\% 
lower limit of 0.3, making this burst not only
an XRF but the most extreme example of an XRF observed so far by
HETE-2.

A comparison of the power-law and Band function fits to the first peak,
which has a much higher S/N than the second peak, provides modest
evidence for an $E^{\rm obs}_{\rm peak}$ near 2 keV, the lower limit of
the energy range of the WXM detectors.  Specifically, we find
that$\Delta \chi^2 = 4.34$ for one additional parameter, which means
that the data requests the (more complicated) Band function model at
the 89\% confidence level.  However, the evidence is clearly not of
high statistical significance, and in this fit we fixed $\alpha$ at $-$1,
its typical value for GRBs.

We therefore choose to place an upper limit on $E^{\rm obs}_{\rm
peak}$.  The appropriate model to use is the Band function, since (as
we have already mentioned) the spectra of almost all GRBs and at least
some XRFs are well-described by this function.  However, this presents
a problem:  the Band function has two distinct ways of representing a
power-law spectrum in the detector energy range.  First, it can do so
by having $E_{\rm break} \rightarrow 0$, so that only the high-energy,
pure power-law part of the Band function is visible in the energy range
of the detector.  Second, it can do so by having $E_{\rm break}
\rightarrow \infty$ and $E_0 \rightarrow \infty$, where $E_0$ is the
``cutoff energy'' of the cutoff power-law that constitutes the
low-energy part of the Band function.  In this limit, the limiting
power-law is actually the cutoff power-law, but the cutoff energy is so
large that the curvature of the model is imperceptible in the detector
energy range.

We solve this problem by developing a new statistical method.  This 
method uses a {\it constrained} Band function which is parameterized by
two quantities, $E^{\rm obs}_{\rm peak}$ and $\beta$.   The {\it
constrained} Band function is perfectly able to make both pure
power-law spectra and power-law times exponential spectra of the
required curvature in the detector energy range, but only the
high-energy part of the Band function is allowed to produce a pure
power-law spectrum.  We describe this new method in detail in Appendix
A.  This method has general applicability to all instruments when the
spectra of the bursts considered have $E^{\rm obs}_{\rm peak}$ near or 
below the low-energy threshold of the detector.  It is necessary 
to demonstrate that the photon index $\beta < -2$  before applying
this method.

In applying the constrained Band function method to XRF 020903,
we jointly fit the WXM and the FREGATE data.  Figure 5 shows the
posterior probability density distribution for $E^{\rm obs}_{\rm peak}$
that we find using this approach.  From this posterior
probability density distribution, we find a best-fit value $E^{\rm
obs}_{\rm peak} = 2.7$ keV,  that $1.1\ {\rm keV} < E^{\rm obs}_{\rm
peak} < 3.6\ {\rm keV}$ with 68\% probabilty, and that $E^{\rm
obs}_{\rm peak} <$ 4.1 and 5.0 keV with 95\% and 99.7\% probabilities,
respectively.

We conclude that the properties of XRF 020903 are very similar to those
of long GRBs, with the exception that the observed peak energy $E^{\rm
obs}_{\rm peak} \sim 3$ keV is $\sim 100$ times smaller.  The extremely
low value of $E_{\rm peak}^{\rm obs}$ seen in XRF 020903 is similar to
the smallest value found among the 9 XRFs whose spectra were determined
by jointly fitting {\it Beppo}SAX WFC and BATSE data \citep{kippen2002}.

\section{Discussion}

\subsection{Source Properties}

We exclude the possibility that XRF 020903 is a Type I X-ray burst
(XRB) on the following grounds.  First, its galactic latitude  is $b =
-61.5^\circ$ (using the center of the combined WXM plus SXC error box),
and there is no known persistent X-ray source or globular cluster in 
this error box.  Since Type I X-ray burst sources lie in the Galactic
plane or in globular clusters, and have persistent X-ray emission, XRF
020903 is unlikely to be  an X-ray burst on locational grounds alone. 
Second, the time history of XRF 020903 is not FRED-like (i.e., it does
not exhibit a fast rise and an  exponential decay), while those of XRBs
typically are.  Third, although the blackbody model gives an acceptable
fit to the spectra of the first and second peaks in the time history of
XRF 020903, the derived blackbody temperatures are $\sim$ 1.0 keV. 
These temperatures are lower than those of almost all Type I X-ray
bursts [which typically have temperatures $T \approx 2$~keV; see,  e.g.
\cite{lewin1993}].  For these reasons, we conclude that XRF 020903 is
an XRF and not an XRB.

The extremely low value of $E^{\rm obs}_{\rm peak}$ observed for XRF
020903 is remarkable.  If the observed spectrum of XRF 020903 were the
redshifted spectrum of a typical GRB, the implied redshift would be $z
\sim 100$, using the best-fit value of $E^{\rm obs}_{\rm peak} = 2.7$
keV observed for XRF 020903 and the mean value of $E^{\rm obs}_{\rm
break}$ for the sample of 5500 spectra formed from the brightest 156
BATSE GRBs \citep{preece2000}.   A redshift of this magnitude would be
hard to understand, and is certainly not expected if long GRBs are
associated with the collapse of massive stars \citep{lamb2000}.  It is
also wildly inconsistent with the measured redshift $z = 0.25$ of the
host galaxy \citep{soderberg2002} of the candidate optical
\citep{soderberg2002} and radio \citep{berger2002} afterglows of XRF
020903. It is therefore difficult to attribute the low observed value
of $E^{\rm obs}_{\rm peak}$ for XRF 020903 to cosmological redshift.

\subsection{Fluence and Peak Energy Correlations}

In Figure 6, we plot XRF 020903 in the ($S_{\rm 30-400}$,$S_{\rm
7-30}$)-plane, where $S_{\rm 7-30}$ and $S_{\rm 30-400}$ are the energy
fluences of the bursts in the 7-30 and 30-400 keV energy bands.  For
the value of $S_{\rm 7-30}$ and $S_{\rm 30-400}$, we use the 
best-fit power-law model for the average spectrum of WXM and FREGATE.  
Also plotted in this figure are the 35 GRBs
whose spectra have been determined using HETE-2 FREGATE data
\citep{celine2003}.  Figure 6 shows that the properties of XRF 020903
are consistent with the relation between $S_{\rm 7-30}$ and $S_{\rm
30-400}$ found by \cite{celine2003}.

In Figure 7, we plot XRF 020903 in the ($S_{\rm 2-400}$,$E_{\rm
peak}^{\rm obs}$)-plane, where $E_{\rm peak}^{\rm obs}$ is the peak of
the observed $\nu F_\nu$ spectrum.  For $E_{\rm peak}^{\rm obs}$, we
plot the 99.7\% upper limit (5.0 keV).  The properties of XRF 020903
are consistent with an extension by two decades of the
hardness-intensity correlation \citep{mallozzi1995,lloyd2000} between
$S_{\rm 30-400}$ and $E_{\rm peak}^{\rm obs}$ found by
\cite{celine2003}.

\cite{amati2002} demonstrated that there is a relation between 
$E_{\rm iso}$ and the burst-averaged value of $E_{\rm peak}$ (i.e.,
$E_{\rm peak}$ for the time-averaged  spectrum of the burst).  Assuming
that the candidate optical and radio afterglows of XRF 020903 are
indeed the afterglows of XRF 020903, and therefore that the redshift of
the underlying host galaxy is the redshift of the XRF, we can calculate
the isotropic-equivalent radiated energy $E_{\rm iso}$ and the upper
limit on the burst-averaged peak energy $E_{\rm peak}$ of the
$\nu F_\nu$ spectrum in the source frame, in the same way as did  
\cite{amati2002}. 
Figure 8 shows that the properties of XRF 020903 are consistent with an
extension by a factor of $\sim 300$ in $E_{\rm iso}$ of the relation
found by \cite{amati2002}. 

Figures 6 - 8 provide evidence that XRFs, ``X-ray-rich GRBs,'' and GRBs
form a continuum, and are therefore the same phenomenon.

\subsection{Constraints on Theoretical Models of XRFs}

A variety of theoretical models have been proposed to explain XRFs
[see, e.g., \cite{zhang2003} for a comparative discussion of several of
these models].  In the off-axis GRB jet model (Yamazaki, Ioka, \&
Nakamura 2002, 2003), XRFs are the result of viewing the jet of an
ordinary GRB off-axis, so that relativistic beaming shifts the
$\gamma$-rays into the X-ray range.  In the clean fireball model, XRFs
are due to the relativistic pair plasma in the GRB jet becoming
optically thin much later than usual, at which time the relativistic
bulk Lorentz factor $\Gamma$ has already decreased to a relatively low
value \citep{mochkovitch2003}.  In the dirty fireball model, XRFs occur
when there is significant baryon loading of the GRB jet, so that
$\Gamma$ never reaches large values \citep{dermer1999,huang2002}.  In
the universal jet model, XRFs are the result of viewing the GRB jet
off-axis, where $\Gamma$ is lower because of the structure of the jet
\citep{rossi2002,woosley2003,zhang2002,meszaros2002}.  In the uniform
jet model, the different properties of XRFs, ``X-ray-rich'' GRBs, and
GRBs are due primarily to different jet opening angles, with larger
jet opening angles associated with lower values of $\Gamma$ 
\citep{lamb2003}.

Any such model of XRFs must reproduce the correlation found by
\cite{celine2003} between $S_{\rm 7-30}$ and $S_{\rm 30-400}$, and the
evidence we report in this paper for correlations between $S_{\rm
2-400}$ and $E_{\rm peak}^{\rm obs}$, and especially, $E_{\rm iso}$
and $E_{\rm peak}$ -- the latter spanning nearly five decades in
$E_{\rm iso}$.

\section{Conclusions}

In this paper, we have reported HETE-2 WXM/FREGATE observations of the X-ray
flash, XRF 020903.  This event was extremely soft: the spectrum had a
best-fit peak energy $E_{\rm peak}^{\rm obs} = 2.7$ keV and  $E_{\rm
peak}^{\rm obs} < 5.0$ keV (99.7\% probability upper limit) and no
photons were detected above $\sim$ 10 keV.  The burst is shorter at
higher energies, which is typical of long GRBs.  We considered the
possibility that the burst lies at very high redshift and that the low
value of $E_{\rm peak}^{\rm obs}$ is therefore due to the cosmological
redshift, and showed that this is very unlikely.  We find that the
properties of XRF 020903 are consistent with the relation between
$S_{7-30}$ and $S_{30-400}$ found by Barraud et al. (2003) for GRBs and
X-ray-rich GRBs, and are consistent with an extension by two decades of
the hardness-intensity correlation \citep{mallozzi1995,lloyd2000}
between $S_{30-400}$ and $E_{\rm peak}^{\rm obs}$ demonstrated by the
same authors.  Assuming that XRF 020903 lies at a redshift $z = 0.25$
as implied by the host galaxy of the candidate optical afterglow of
this burst, we find that the the properties of XRF 020903 are
consistent with an extension by a factor $\sim 300$ of the relation
between $E_{\rm iso}$ and $E_{\rm peak}$ in the source frame of the
burst found by \citet{amati2002} for GRBs.  When combined with earlier
results, the results reported in this paper provide strong evidence
that XRFs, X-ray-rich GRBs, and GRBs form a continuum and are a single
phenomenon.  The correlation found by \cite{celine2003} between $S_{\rm
7-30}$ and $S_{\rm 30-400}$, and the evidence we find in this paper for
correlations between $S_{30-400}$ and $E_{\rm peak}^{\rm obs}$, and
especially, $E_{\rm iso}$ and $E_{\rm peak}$, provide strong
constraints on any model of XRFs and X-ray-rich GRBs. 






\acknowledgments

We would like to thank the anonymous referee for comments and
suggestions that materially improved the paper.  The HETE-2 mission is
supported in the U.S. by NASA contract NASW-4690; in Japan, in part by
the Ministry of Education, Culture, Sports, Science, and Technology
Grant-in-Aid 12440063; and in France, by CNES contract 793-01-8479.  K.
Hurley is grateful for {\it Ulysses} support under contract JPL 958059
and for HETE-2 support under contract MIT-SC-R-293291.  G. Pizzichini
acknowledges support by the Italian Space Agency.  One of the authors
(TS) is partially supported by the Junior Research Associate (JRA)
program at RIKEN.  



\appendix
\section{The ``Constrained'' Band Function For Soft GRBs}

In the spectral analysis of GRBs, one occasionally encounters events
(such as XRF 020903) that are so soft that they present themselves as
pure power-laws with power-law index $\beta < -2$ in the energy range
of the detector.  The natural interpretation of such spectra is that
the break energy $E_{\rm break}$ separating the two functional parts of
the Band function is near or below the lower boundary of the detector
energy range.

This situation creates a problem for fits of the Band function, in that
the Band function has two distinct ways of conforming to a power-law in
the detector energy range:

\begin{enumerate}

\item 
$E_{\rm break}\rightarrow 0$, so that only the high-energy, pure
power-law part of the Band function is visible in the energy range of
the detector.

\item 
$E_{\rm break}\rightarrow \infty$, $E_0\rightarrow \infty$, where $E_0$
is the ``cutoff energy'' of the cutoff power-law that constitutes the
low-energy part of the Band function.  In this limit, the limiting
power-law is actually the cutoff power-law, but the cutoff energy is so
large that the curvature of the model is imperceptible in the detector
energy range.

\end{enumerate}

Therefore, despite the fact that the numerical value of the power-law
index is such that we are certain that we should be dealing with the
high-energy part of the Band function (i.e., the index is $< -2$), the
low-energy part of the function can ``horn in'' on the fit, altering 
the physical inferences drawn from the spectrum.

This situation is particularly a problem for the estimation of $E_{\rm
peak}$.  Since we know that we are in case 1, we also know that we
ought to have at least a firm upper limit on $E_{\rm peak}$, since
$E_{\rm peak}$ is always necessarily less than $E_{\rm break}$, which
is at the low end of the detector energy range.   On the other hand,
the case 2 limit implies $E_{\rm peak}\rightarrow \infty$. 
Unfortunately, the data don't care which side of the Band function
makes the power-law, so no discrimination is possible between the two
cases.  Consequently, we can't constrain $E_{\rm peak}$ at all using a
normal Band function fit.

The approach we have chosen to deal with this situation in the case of
XRF 020903 is to fit a {\it constrained} Band function to
the data.  That is, we consider a three-dimensional subspace of the
full four-dimensional Band function parameter space, choosing the
subspace with a view to satisfying the following criteria:

\begin{enumerate}

\item 
It is perfectly possible to make both pure power-laws and cutoff
power-laws of the any desired curvature in the detector energy range.

\item 
Only the high-energy part of the Band function is allowed to produce a
pure power-law.

\end{enumerate}

We define the three-dimensional subspace in the following way: 
consider a Band function parametrized by low- and high-energy indices
$\alpha$ and $\beta$, and by a cutoff energy $E_0$.  The well-known
relation between $E_0$ and $E_{\rm break}$ is $E_{\rm break} =
(\alpha-\beta) E_0$.  We impose the constraint condition on our family
of fitting functions
\begin{equation}
E_{\rm break}=E_{\rm pivot}\times(E_0/E_{\rm pivot})^{-1},
\end{equation}
where $E_{\rm pivot}$ is some suitably chosen energy, in the general
neighborhood where the GRB has appreciable emission.  $E_{\rm break}$
and $E_0$ are then inversely related, and are equal to each other when
both are equal to $E_{\rm pivot}$.

When $E_0 < E_{\rm pivot}$, then $E_{\rm break} > E_{\rm pivot}$, and
the function is essentially a cutoff power-law in the energy range of
interest.

On the other hand, when $E_0  > E_{\rm pivot}$, then $E_{\rm break} <
E_{\rm pivot}$, and as $E_0 \rightarrow \infty$, $E_{\rm break}
\rightarrow 0$.  In other words, when the low-energy part of the Band
function is trying to imitate a power law, the break energy becomes
small enough to force the low-energy part of the function below the
energy range of interest, where it cannot be seen and therefore can do
no harm.  Any pure power-law work must thus be done by the high-energy
part of the Band function.

The resulting spectral function has three parameters
(including the scale), rather than four.  The two input shape
parameters can be chosen arbitrarily from the set \{$\alpha$, $\beta$,
$E_0$, $E_{\rm break}$, $E_{\rm peak}$\}.  The remaining parameters may
then be determined by algebraic relationships. 

We have found it most convenient to adopt $E_{\rm peak}$ and $\beta$ as
our parameters.  The choice of $E_{\rm peak}$ is dictated by the
necessity of estimating its value, or at least an upper bound on its
value.  The choice of $\beta$ is convenient because one may then impose
the parameter bound $\beta < -2$, which guarantees that the formal
expression for $E_{\rm peak}$ may be meaningfully interpreted as the
energy of the peak of the $\nu F_\nu$ distribution.  This bound on
$\beta$ is an important part of the specification of the fitting family
of models.  Were it not imposed, it would be possible for the formal
expression for $E_{\rm peak}$ to exceed $E_{\rm break}$, so that at
large values of $E_{\rm peak}$ the fit could always produce a
$\beta\gtrsim -2$ power-law in the detector energy range.  The result
would be an extended tail of constant $\chi^2$ for arbitrarily large
values of $E_{\rm peak}$.

Figure 9 shows the constrained Band function, with $\beta=-2.5$ and
$E_{\rm pivot}=4$~keV, for different values of $E_{\rm peak}$.  This
figure shows that $E_{\rm peak}$ increases, $E_{\rm break}$ also necessarily
increases, so that $E_0$ is forced to smaller and smaller values by the
constraint, which increases the curvature (and the value of $\alpha$).

Figure 10 shows the constrained Band function, with $E_{\rm peak}=4$~keV
and $E_{\rm pivot}=4$~keV, for different values of $\beta$.  The
progression from some curvature at low energy ($\beta=-2.0$) to almost
none ($\beta=-4.0$) is evident, as is the fact that as the curvature
disappears, the resulting power-law is produced by the high-energy part
of the Band function.

Figures 9 and 10 show that the constrained Band function is perfectly
able to make both pure power-laws, and cutoff power-laws with any
desired curvature in the detector energy range.  Figure 10 demonstrates
that, in the constrained Band function, a power-law spectrum is always
produced by the high-energy part of the Band function.

The choice of $E_{\rm pivot}$ is dictated by the following
considerations:

\begin{enumerate}

\item 
$E_{\rm pivot}$ must be low enough to prevent the low-energy part of
the Band function from making a power-law in the energy range of
interest.  If $E_{\rm pivot}$ were 1 GeV (say), then the Band function
would have no difficulty making $E_0$ large and $\alpha\lesssim -2$,
which is what we are trying to prevent by introducing the constraint. 
So $E_{\rm pivot}$ should be ``as low as possible.''

\item 
$E_{\rm pivot}$ must not be so low that we cannot adequately fit any
curvature that may exist in the spectrum.  If $E_{\rm pivot}$ were 1~eV
(say), then whenever $E_{\rm break}$ was in or above the energy range
where the spectrum is appreciable, $E_0$ would be so tiny that the
curvature of the model would be huge, much too large to fit the data
well.

\end{enumerate}

One way of choosing $E_{\rm pivot}$ is to calculate its value using
the best-fit parameters from a fit of a {\it free} Band function, using
$E_{\rm pivot}=(E_0E_{\rm break})^{1/2}$.  This choice, which
effectively chooses the unique constrained subspace of the full
parameter space that contains the best-fit free Band function, allows
the constrained family of functions to optimally fit whatever curvature
the data may seem to hint is required.

We must require that the inferences that we draw from the spectral fit
should be robust, in the sense that they should not depend strongly on
the specific choice of $E_{\rm pivot}$.  So the proper use of this
constrained Band model involves not only choosing a representative
value of $E_{\rm pivot}$, but also varying $E_{\rm pivot}$ in some
reasonable range, to make sure that the conclusions about parameter
estimates and bounds are unaffected by the choice of $E_{\rm pivot}$.

Figure 11 shows the constrained Band function, with $E_{\rm peak}=4$~keV
and $\beta=2.5$, for different values of $E_{\rm pivot}$.  Once again, as
the low-energy curvature disappears, the resulting power-law is
produced by the high-energy part of the Band function.  Figure 11 also
shows that the shape of the spectrum in the detector energy range is
insensitive to the specific choice of $E_{\rm pivot}$, within a
reasonable range.  Thus the conclusions about parameter estimates and
bounds are unaffected by the choice of $E_{\rm pivot}$.

Figure 12 shows the constrained Band functions with parameters that
best fit the 13~s spectrum of XRF 020903, for different fixed values of
$E_{\rm pivot}$.  This figure illustrates the fact that the shape of the
best-fit model is essentially unchanged in the energy range of the WXM
for choices of $E_{\rm pivot}$ within a reasonable range.`

Finally, we give the algebraic relationships necessary to recover the
remaining Band function parameters assuming that $E_{\rm peak}$ and
$\beta$ are given.  Let $x\equiv E_{\rm peak}/E_{\rm pivot}$.  Then, %
\begin{eqnarray}
\alpha &=& -2 + \frac{1}{2}x^2 + 
           \sqrt{\frac{1}{4}x^4 - x^2(\beta + 2)},\label{alpha}\\
E_0 &=& (2 + \alpha)E_{\rm peak}.
\end{eqnarray} 
In Equation (\ref{alpha}), we have resolved the ambiguity in the choice of
root of a quadratic equation by requiring that when $\beta+2 < 0$,
then  $\alpha+2 > 0$, so that $E_{\rm peak}$ is in fact the peak energy
of the  $\nu F_\nu$ distribution.




\clearpage

\begin{table}
\begin{center}
\caption{Temporal properties of XRF 020903.}
\vspace{5mm}
\begin{tabular}{ccc}\hline\hline
Energy Band & t$_{50}$     & t$_{90}$    \\
   (keV)    &  (s)         &   (s)       \\\hline
 2 - 5      & 5.8$\pm$0.9  & 10.6$\pm$0.2 \\
 5 - 10     & 2.4$\pm$0.2  &  4.3$\pm$2.2 \\
 2 - 10     & 4.9$\pm$0.6  &  9.8$\pm$0.6 \\\hline
\end{tabular}
\end{center}
\centerline{Note.---The quoted errors correspond to $\pm 1\sigma$.}
\end{table}

\begin{deluxetable}{ccccccc}
\tabletypesize{\small}
\tablewidth{0pt}
\tablecaption{Results of fits to the spectrum of XRF 020903.}
\startdata\hline\hline
Time region & Model & kT & $\alpha$ & $\beta$ & 
$E_{\rm peak}^{\rm obs}$ & $\chi^{2}_{\nu}$ (DOF) \\
(s)      &       & (keV) & & & (keV) \\\hline
0.0-8.0  & blackbody & 1.04$_{-0.20}^{+0.24}$ & & & & 1.08 (62)\\
         & power-law & & $-$2.4$_{-0.6}^{+0.5}$ & & & 1.21 (62)\\
         & cutoff power-law & & $-$1.0 (fixed) &  
	 & 3.1$_{-1.1}^{+1.9}$ & 1.14 (62)\\
         & Band & & $-$1.0 (fixed) & 
    $<$ $-$2.4 & 3.4$_{-1.0}^{+1.7}$ & 1.16 (61)\\
8.0-13.0 & blackbody & 0.54$_{-0.23}^{+0.23}$ & & 
 & &  1.13 (62)\\
         & power-law & & $-$4.2$_{-3.7}^{+1.1}$ & & & 1.15 (62)\\
         & cutoff power-law & & $-$1.0 (fixed) & & 
    $<$ 2.0 & 1.14 (62)\\
         & Band && $-$1.0 (fixed) 
  & $<$ $-$3.3 & $<$ 2.0 & 1.15 (61)\\
0.0-13.0 & blackbody & 0.87$_{-0.16}^{+0.20}$ & &
   & & 0.79 (62)\\
         & power-law & & $-$2.8$_{-0.6}^{+0.5}$ & 
    & & 0.86 (62)\\
         & cutoff power-law & & $-$1.0 (fixed) & 
    & 2.4$_{-0.7}^{+1.2}$ & 0.81 (62)\\
         & Band & & $-$1.0 (fixed)
  & $<$ $-$2.7 & 2.4$_{-0.7}^{+1.2}$ & 0.82 (61)\\
0.0-13.0 & blackbody & 0.90$_{-0.17}^{+0.21}$ & & 
    & & 0.85 (177)\\
  (with FREGATE) & power-law & & $-$2.6$_{-0.5}^{+0.4}$ &
  & & 0.86 (177)\\
         & cutoff power-law & & $-$1.0 (fixed) & 
  & 2.6$_{-0.8}^{+1.4}$ & 0.85 (177)\\
         & Band & & $-$1.0 (fixed) 
	 & $<$ $-$2.3 & $<$ 4.1 & 0.86 (176)\\
\enddata

{Note.---The quoted errors correspond to the 90\% confidence region}

\end{deluxetable}

  \begin{table}
  \begin{center}
  \caption{Peak photon number and energy fluxes (in 1 s) and fluences in
  various energy bands for XRF 020903.}
  \vspace{5mm}
  \begin{tabular}{lccc}\hline\hline
            & 2-5 keV & 5-10 keV & 2-10 keV 
  \\
  \hline
  Peak flux (ph cm$^{-2}$ s$^{-1}$) & $1.9 \pm 0.7$ & 
  0.33$_{-0.16}^{+0.19}$ & $2.2 \pm 0.8$
  \\
  Peak flux (10$^{-9}$ ergs cm$^{-2}$ s$^{-1}$) & $10.4_{-3.7}^{+3.6}$ &
  $4.3_{-2.2}^{+2.3}$ & $14.7 \pm 5.3$\\
  Total fluence (10$^{-8}$ ergs cm$^{-2}$) & $4.2 \pm 0.9$ &
  1.7$_{-0.7}^{+0.8}$ & $5.9 \pm 1.4$\\\hline
  \end{tabular}
  \end{center}
  {\qquad \qquad \quad Note.---All of the quantities in this table are
  derived assuming a power-law model for the spectrum.  The quoted
  errors correspond to the 90\% confidence region.}
  \end{table}

\begin{figure}
\includegraphics[width=6.0truein,clip=]{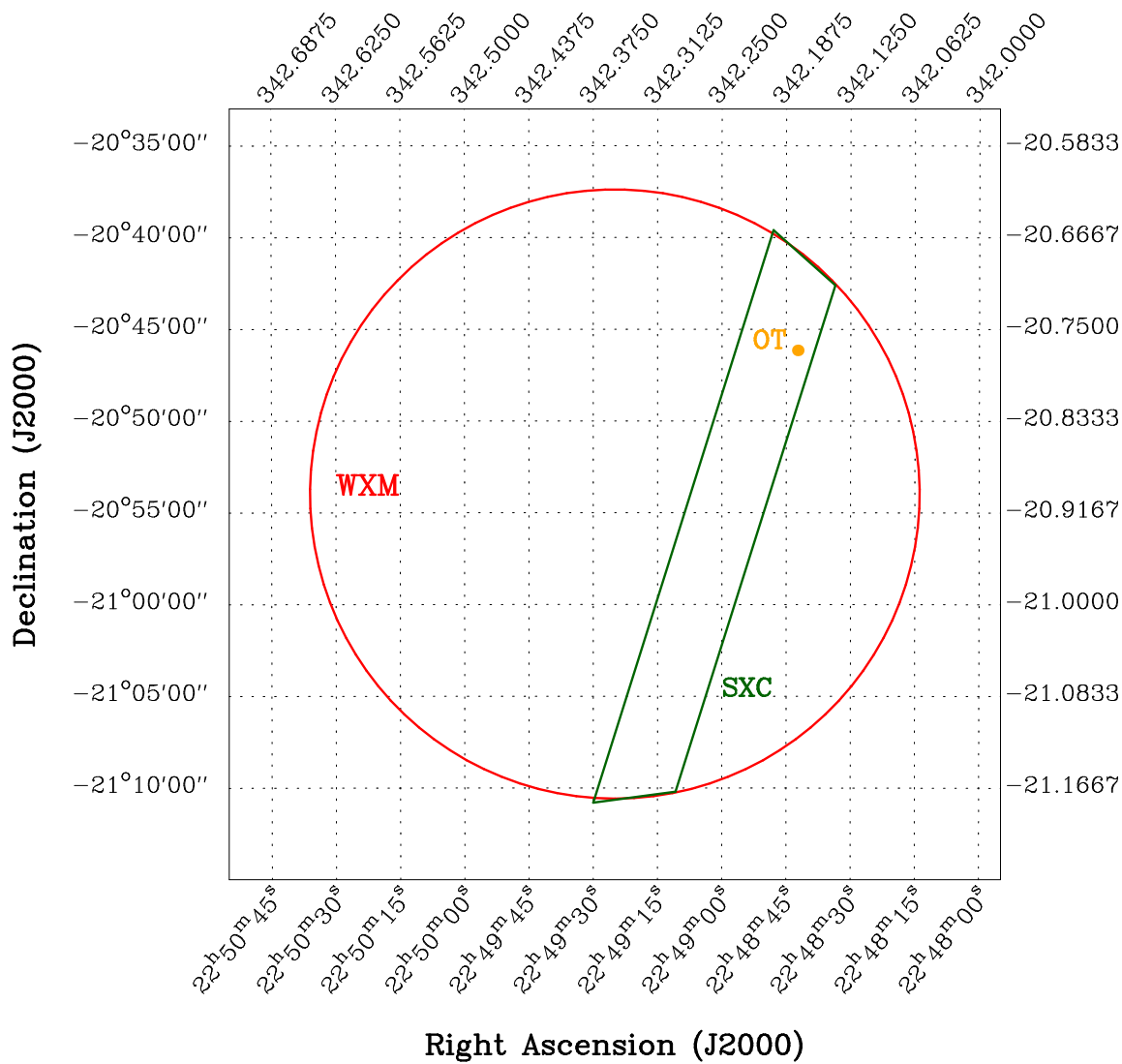}
\caption{The HETE-2 WXM/SXC localization for XRF 020903.  The circle is
the 90\% confidence region for the WXM localization and the belt-like 
region is the portion of the 90\% confidence region for the
one-dimensional SXC localization that lies within the WXM 90\%
confidence circle.   The final localization is the intersection of the
WXM and SXC localizations (Ricker et al. 2002a).  The point labeled
``OT'' is the location of the candidate optical (Soderberg et al. 2002)
and radio (Berger et al. 2002) afterglows of XRF 020903.  
\label{fig1}}
\end{figure}

\begin{figure}
\includegraphics[width=6.0truein,clip=]{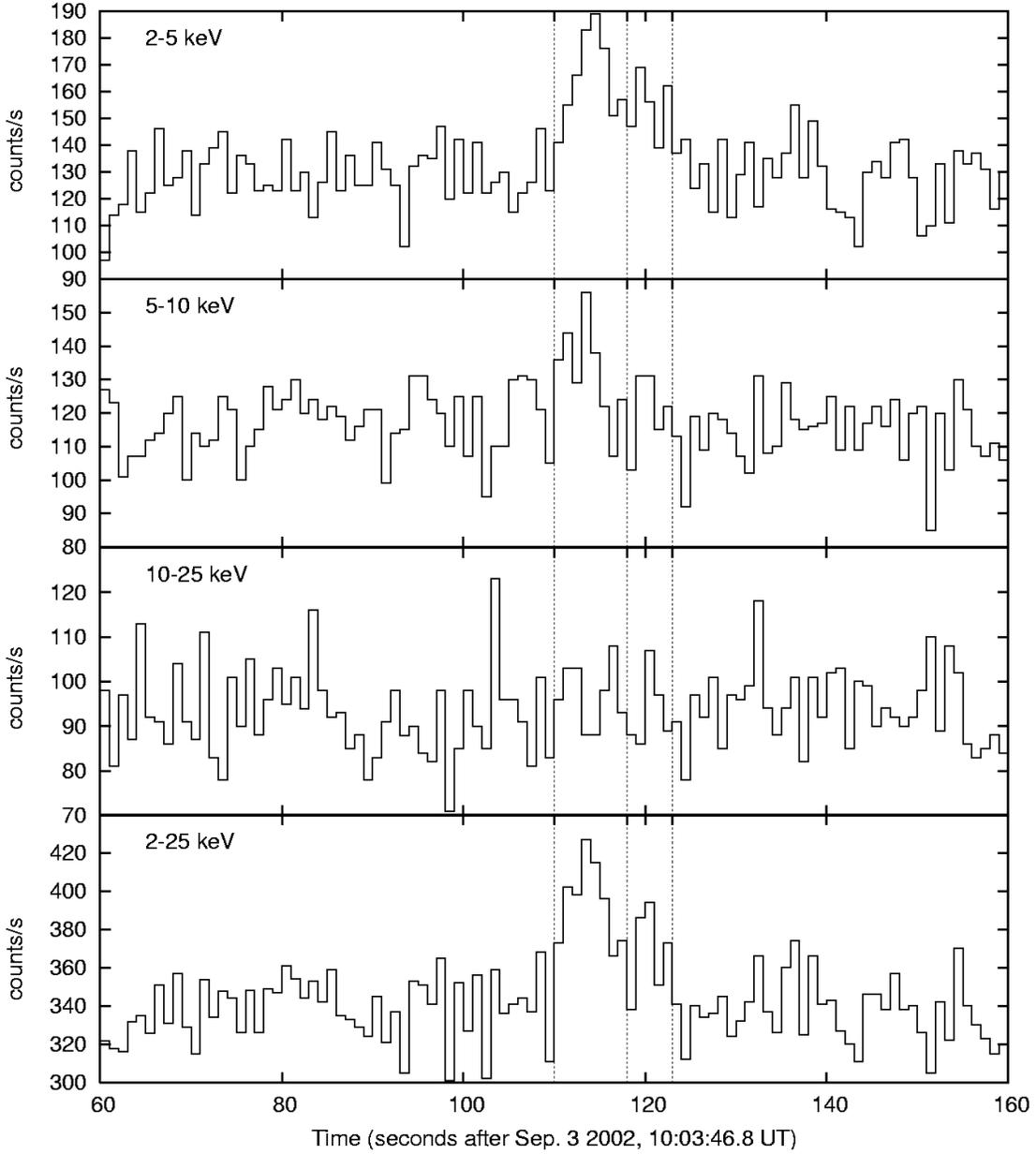}
\caption{The light curve of XRF 020903 in four WXM energy bands: 2-5
keV, 5-10 keV, 10-25 keV, and 2-25 keV (top to bottom).  The light
curve is binned in one second bins.  The vertical dotted lines show the
0 - 8 and 8 - 13 second time intervals bracketing the first and second
peaks of the burst light curve.  We have performed  model fits to the
spectra of the burst during these two time intervals, and to the entire
duration of the burst (0 -13 seconds).
\label{fig2}}
\end{figure}

\begin{figure}
\includegraphics[width=4.35truein,angle=270,clip=]{f3a.eps}
\includegraphics[width=4.35truein,angle=270,clip=]{f3b.eps}
\caption{Comparison of the WXM spectra for the time intervals 0-8 and
8-13 s.  The observed (crosses) and predicted (histogram) count rates
are shown in a different color for each of the nine WXM wires that we
have included in the fits.  The spectral model is a power law with
fixed photoelectric absorption (see Table 2).
\label{fig3}}
\end{figure}

\begin{figure}
\includegraphics[width=4.35truein,angle=270,clip=]{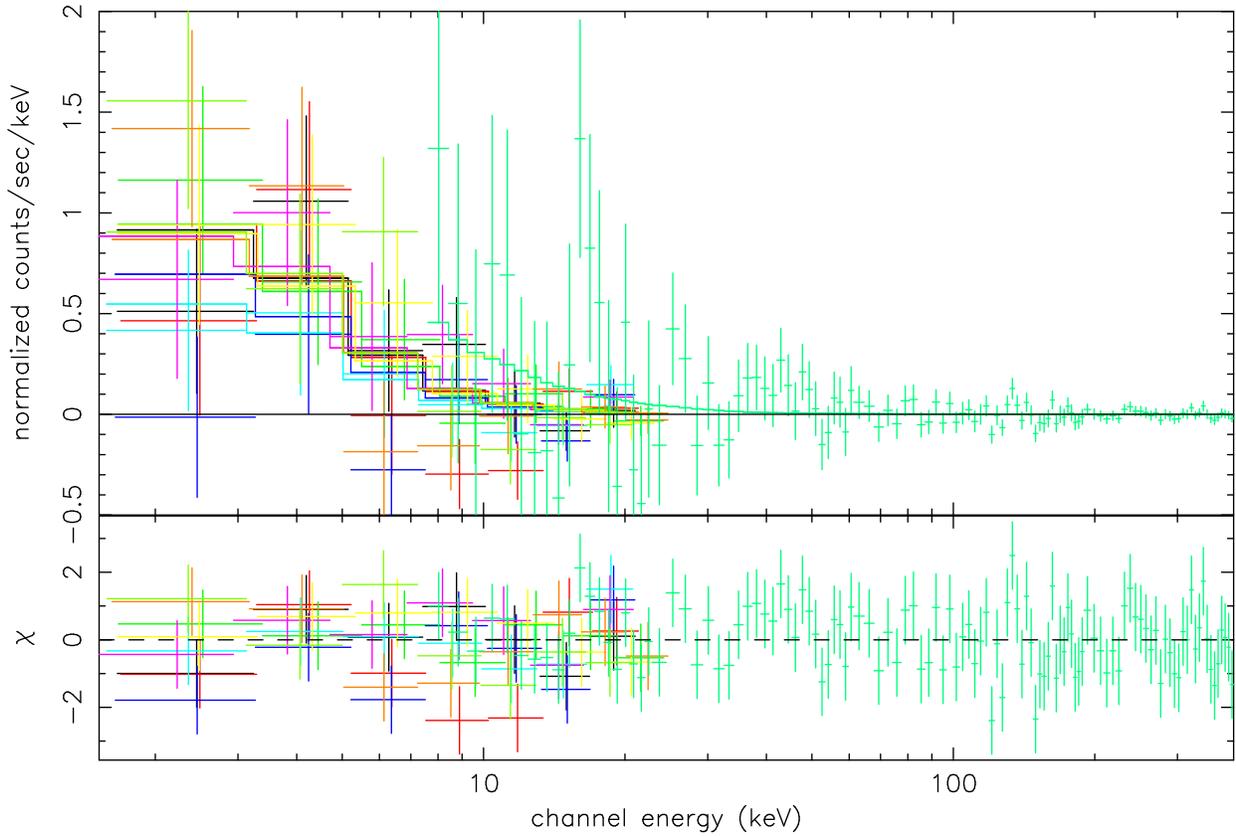}
\caption{The WXM and FREGATE spectra for the entire time interval  0-13
s.  The observed (crosses) and predicted (histogram) count rates are
shown in a different color for each of the nine WXM wires that we have
included in the fits.  The spectral model is a power law with fixed
photoelectric absorption (see Table 2).
\label{fig4}}
\end{figure}

\begin{figure}
\epsscale{0.8}
\includegraphics[width=6.5truein,clip=]{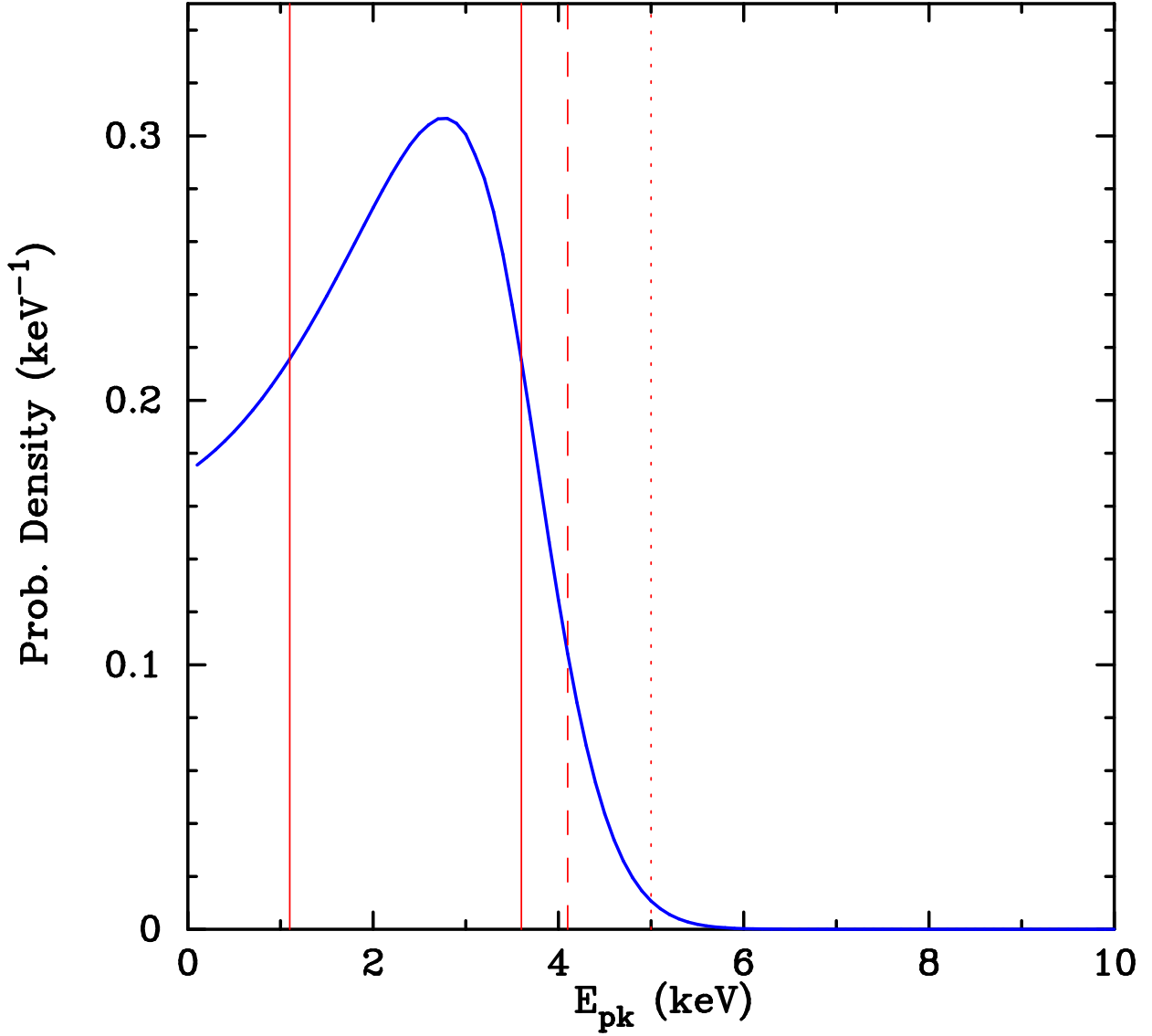}
\caption{The posterior probability density distribution for $E^{\rm
obs}_{\rm peak}$.  The vertical solid lines define the 68\% probability
interval for $E^{\rm obs}_{\rm peak}$, while the dashed and dotted
lines show the 95\% and 99.7\% probability upper limits on $E^{\rm
obs}_{\rm peak}$.  We find a best-fit value $E^{\rm obs}_{\rm peak} =
2.7$ keV, that $1.1\ {\rm keV} < E^{\rm obs}_{\rm peak} < 3.6\ {\rm keV}$
with 68\% probability, and that $E^{\rm obs}_{\rm peak} <$ 4.1
and 5.0 keV with 95\% and 99.7\% probability.
\label{fig5}}
\end{figure}

\begin{figure}
\epsscale{1.0}
\includegraphics[width=4.7truein,angle=270,clip=]{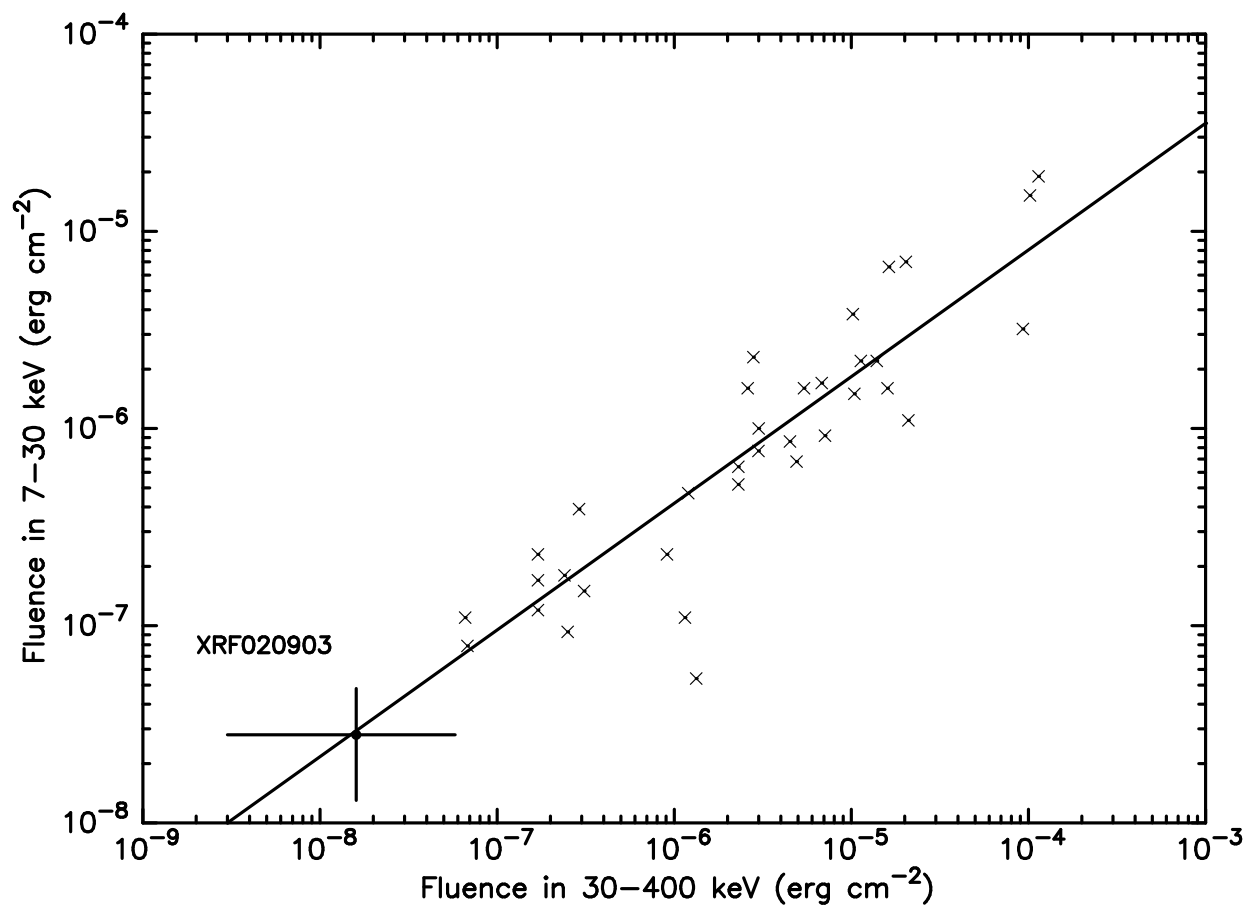}
\caption{The ($S_{7-30}$,$S_{30-400}$)-plane, showing the location of
XRF 020903, using the total fluence in the 7-30 keV
energy band and in the 30-400 keV energy
band (filled circle).  The crosses are the locations of the 35
HETE/FREGATE GRBs studied by \cite{celine2003}.  The solid line is the
relation, $S_{7-30}$ = 3 $\times$ 10$^{-3}$ S$^{0.643}_{30-400}$, found
by \cite{celine2003}.  The properties of XRF 020903 are consistent with
this relation. \label{fig6}}
\end{figure}

\clearpage 
\begin{figure}
\includegraphics[width=4.5truein,angle=270,clip=]{f7.eps}
\caption{The ($S_{2-400}$,$E_{\rm peak}^{\rm obs}$)-plane, showing the
location of XRF 020903.  For $E_{\rm peak}^{\rm obs}$, we plot the
99.7\% probability upper limit (5.0 keV).  We calculate the 2-400
keV fluence using the best power-law model fit jointly to  the WXM and
the FREGATE data.  The crosses show the locations of 12 of the HETE-2
GRBs studied by \cite{celine2003} for which $E_{\rm peak}^{\rm obs}$ is
relatively well determined.  The properties of XRF 020903 are
consistent with an extension by two decades of the hardness-intensity
correlation found by \cite{celine2003}.
\label{fig7}}
\end{figure}

\begin{figure}
\includegraphics[width=6.5truein,clip=]{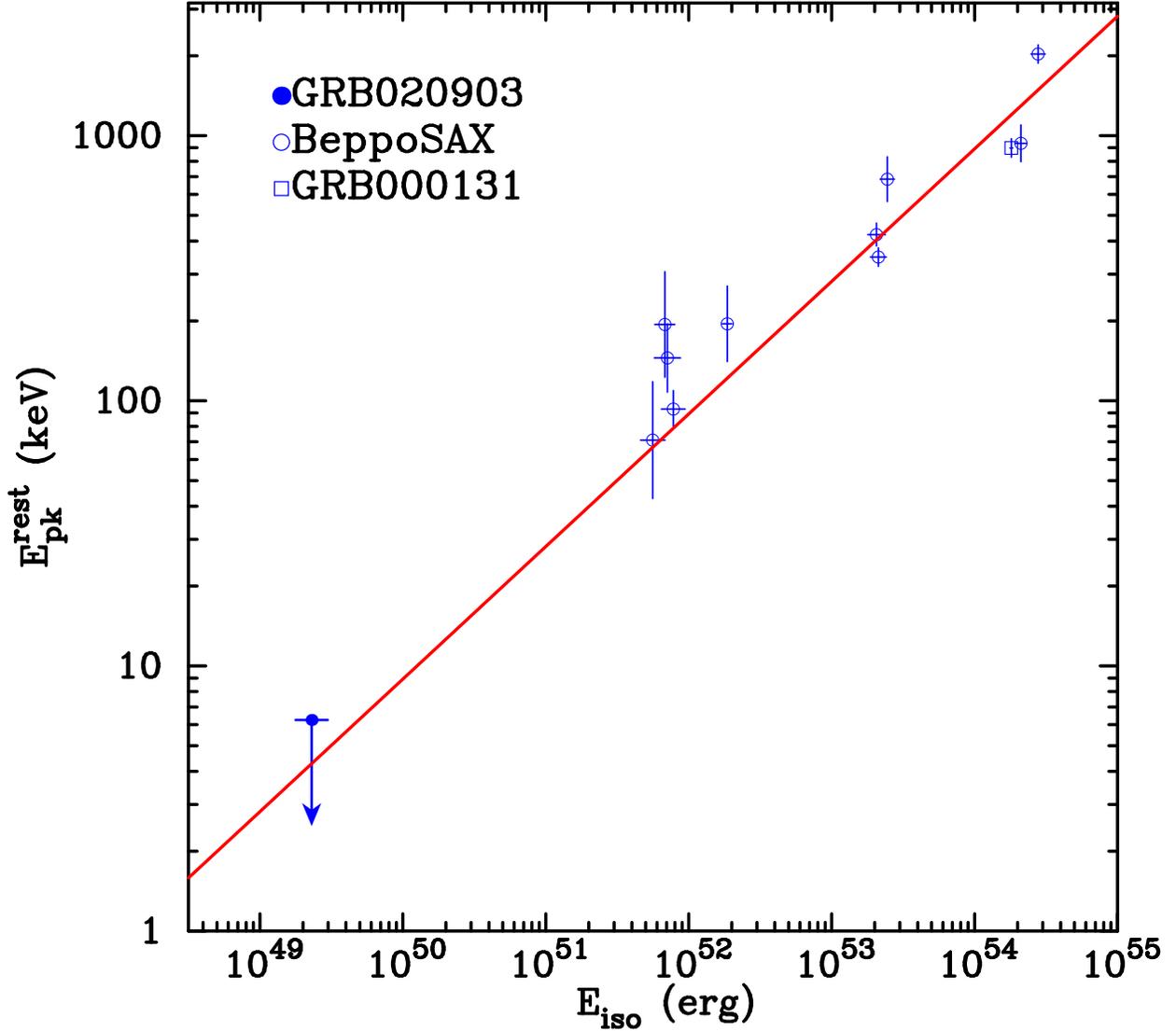}
\caption{The ($E_{\rm iso}$,$E_{\rm peak}$)-plane, where $E_{\rm iso}$
is the isotropic-equivalent radiated energy between 1-10$^{4}$ keV and
$E_{\rm peak}$ is the peak of the $\nu F_\nu$ spectrum, both measured
in the rest frame of the burst.  The filled circle in the lower
left-hand corner is the location of XRF 020903.  The ten open circles
are the {\it Beppo}SAX GRBs reported by Amati et al. (2002).  The solid
line is given by the equation, $E_{\rm peak} = 89 
(E_{\rm iso}/10^{52}{\rm erg})^{0.5}$ {\rm keV}.  The properties of XRF
020903 are consistent with an extension of this relation by a factor
$\sim 300$.
\label{fig8}}
\end{figure}
\clearpage

\begin{figure}
\includegraphics[width=6.5truein,clip=]{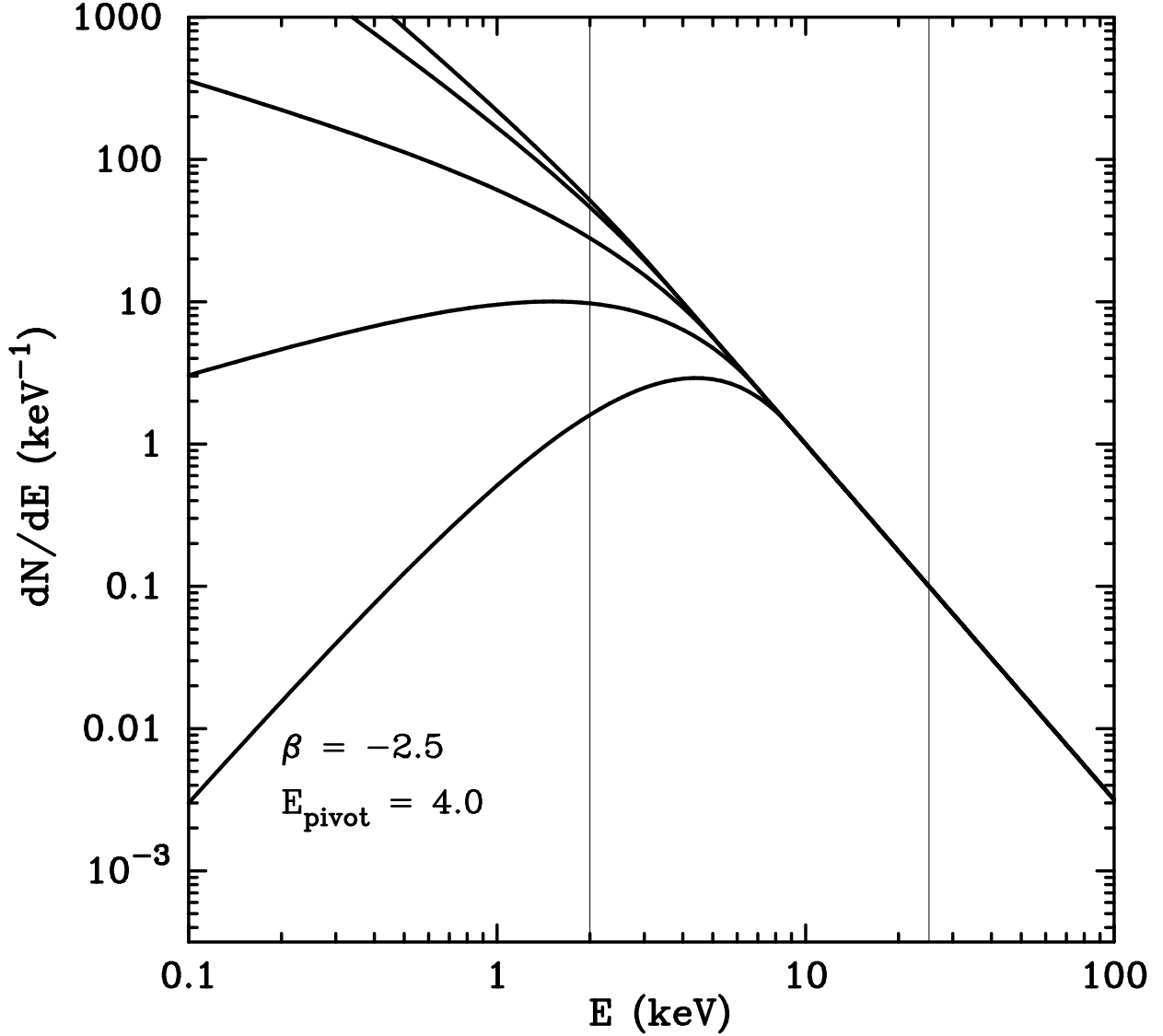}
\caption{Constrained Band functions, with $\beta=-2.5$ and $E_{\rm
pivot}=4$~keV, for different values of $E_{\rm peak}$.  All functions
have been normalized to 1~keV$^{-1}$ at 10~keV.  The two vertical lines
at 2~keV and at 25~keV show the WXM bandpass.  The spectra shown are
(decreasing monotonically from the top at low energy), $E_{\rm
peak}=$1~keV, 2~keV, 4~keV, 6~keV, and 8~keV, respectively.  As $E_{\rm
peak}$ increases, $E_{\rm break}$ also necessarily increases, so that
$E_0$ is forced to smaller and smaller values by the constraint,
increasing the curvature and the value of $\alpha$.
\label{fig9}}
\end{figure}
\clearpage

\begin{figure}
\includegraphics[width=6.5truein,clip=]{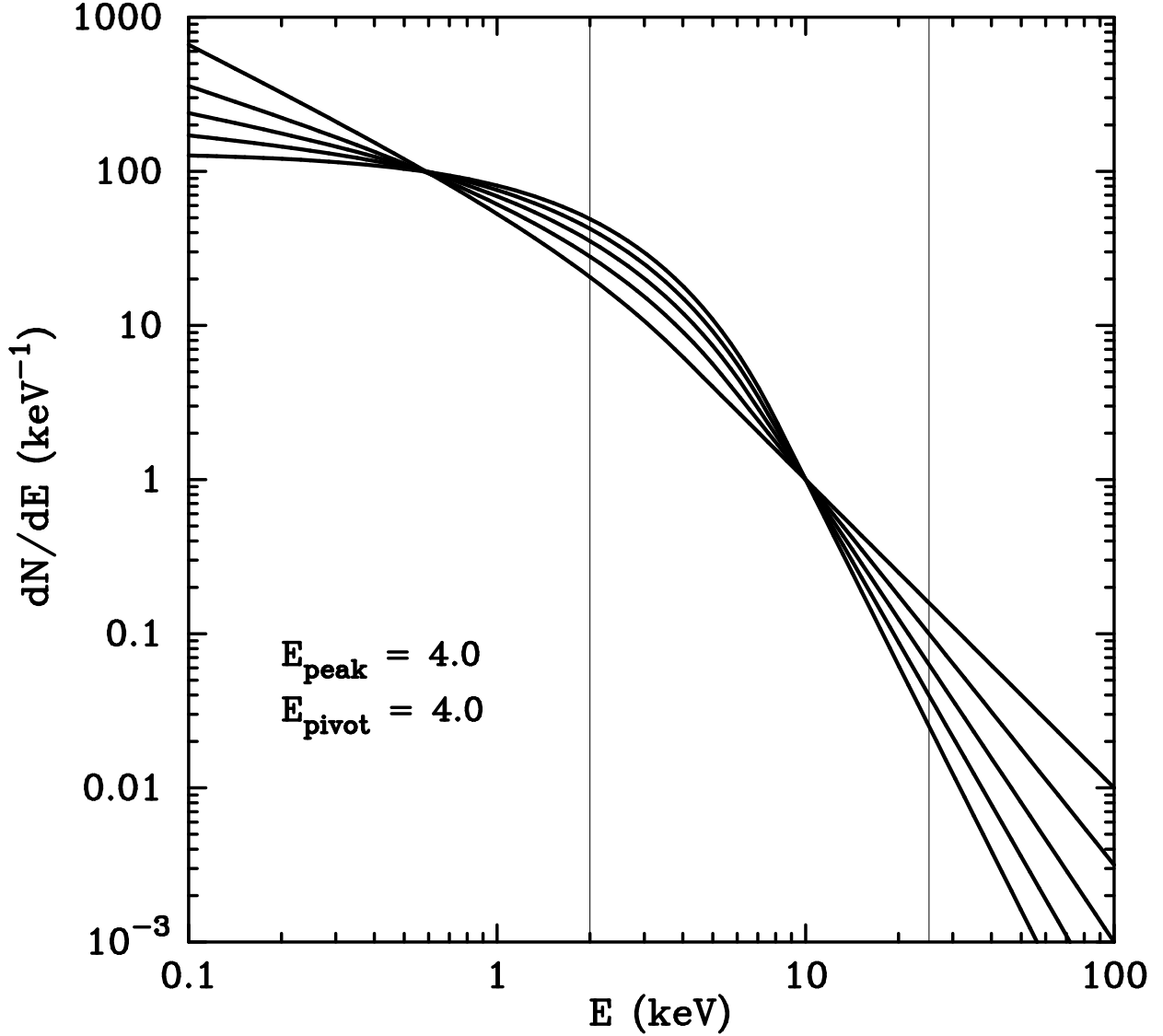}
\caption{Constrained Band functions, with $E_{\rm peak}= 4$~keV and
$E_{\rm pivot}= 4$~keV, for different values of $\beta$.  All functions
have been normalized to 1~keV$^{-1}$ at 10~keV.  The two vertical lines
at 2~keV and at 25~keV show the WXM bandpass.  The spectra shown are
for $\beta =$  -2.0, -2.5, -3.0, -3.5, and -4.0, which can be
distinguished by the increasing steepness of their slopes at high
energy.  The progression from some curvature at low energy
($\beta=-4.0$) to almost none ($\beta=-2.0$) is evident, as is the fact
that as the curvature disappears, the resulting power-law is produced
by the high-energy part of the Band function.
\label{fig10}}
\end{figure}
\clearpage

\begin{figure}
\includegraphics[width=6.5truein,clip=]{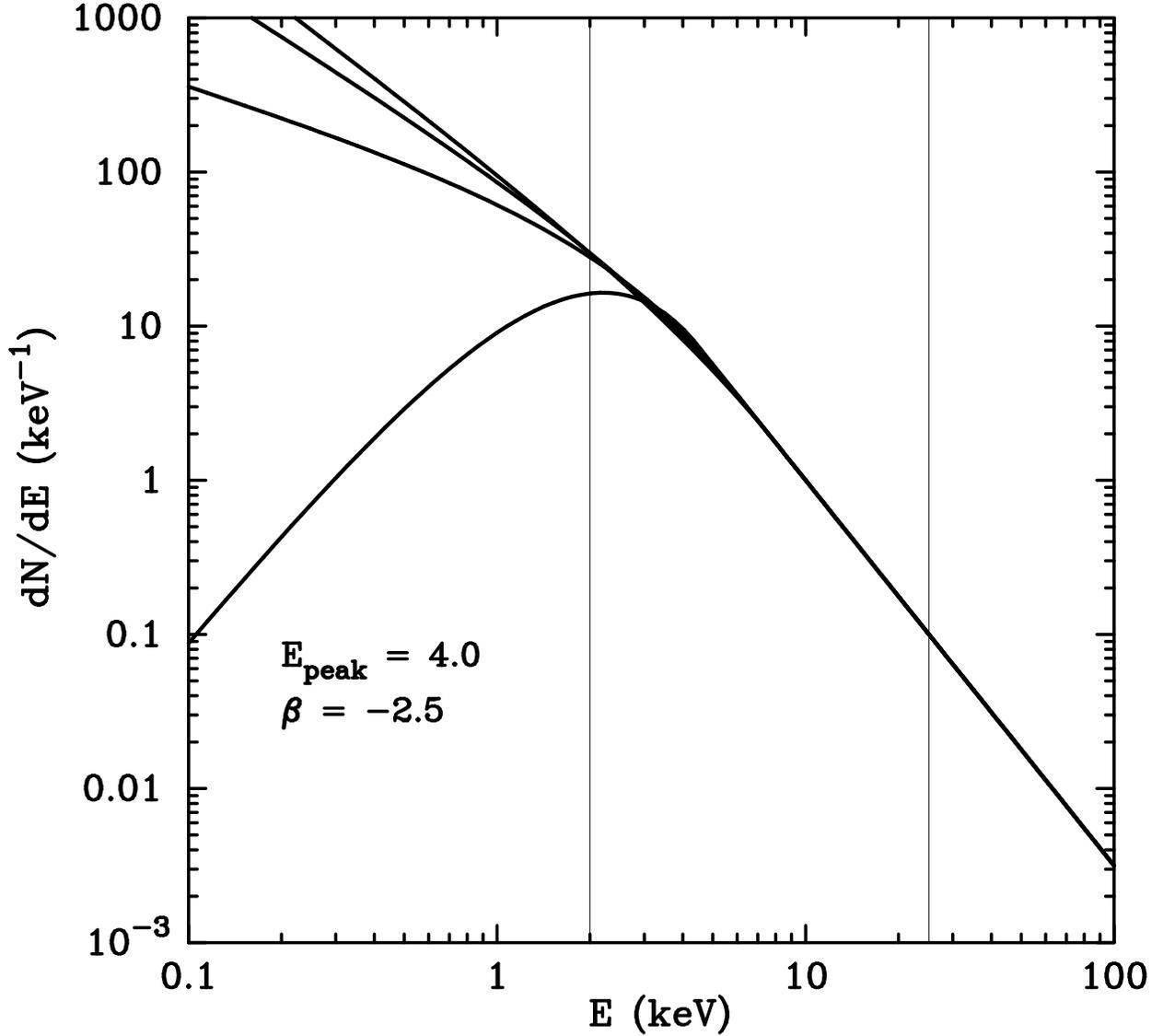}
\caption{Constrained Band functions, with $E_{\rm peak}= 4$~keV and
$\beta=-2.5$, for different values of $E_{\rm pivot}$.  All functions
have been normalized to 1~keV$^{-1}$ at 10~keV.  The two vertical lines
at 2~keV and at 25~keV show the WXM bandpass.  The spectra shown are
(increasing monotonically at low energy) for $E_{\rm pivot}=$\ 2~keV,
4~keV, 6~keV, and 8~keV, respectively.  Once again, as the low-energy
curvature disappears, the resulting power-law is produced by the
high-energy part of the Band function.  Note also that the shape of the
constrained Band function is insensitive to the specific choice of
$E_{\rm pivot}$ within a reasonable range.
\label{fig11}}
\end{figure}
\clearpage

\begin{figure}
\includegraphics[width=6.5truein,clip=]{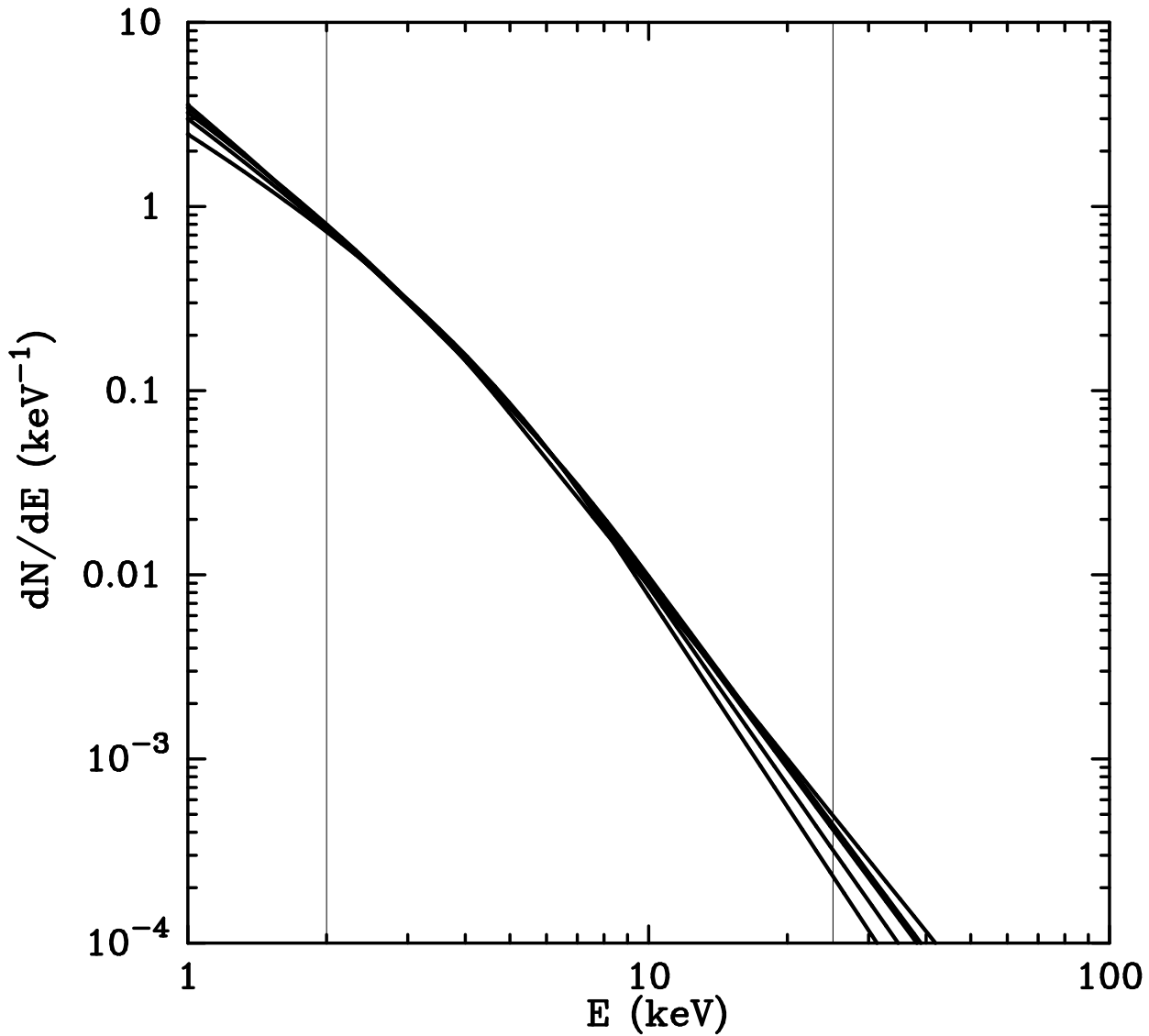}
\caption{
Constrained Band functions with parameters that best fit the 13~s
spectrum of XRF020903, for different fixed values of $E_{\rm pivot}$. 
The two vertical lines at 2~keV and at 25~keV show the WXM bandpass. 
All functions have been normalized so that the integral from 2~keV to
25~keV is one photon. The five spectra shown in the plot corresponding
to $E_{\rm pivot}=$\ 4~keV, 5~keV, 6~keV, 7~keV, and 8~keV (the 7~keV and
8~keV largely overlap each other).  This figure illustrates a robust
aspect of the constraint procedure: the best-fit model is essentially
unchanged in the WXM spectral band despite a factor-of-two change in
the value of $E_{\rm pivot}$.
\label{fig12}}
\end{figure}
\clearpage





\end{document}